\documentclass[technote,onecolumn]{IEEEtran} 

\usepackage{balance}
\usepackage[utf8]{inputenc}
\usepackage[greek,english]{babel}

\IEEEoverridecommandlockouts 
\usepackage[dvips]{graphicx}
\usepackage{algorithmic}
\usepackage{algorithm}
\usepackage{listings}
\usepackage[cmex10]{amsmath}
\usepackage{url}
\usepackage{multirow}
\usepackage{pgfplots} 

\usepackage{amssymb}
\usepackage{amsthm}
\newtheorem{definition}{Definition}

\usepackage{amsthm,amsmath}
\usepackage[utf8]{inputenc} 

\usepackage{amsthm}

\usepackage{subcaption}

\usepackage{amsmath}
\usepackage{amssymb}

\usepackage{pythonhighlight}

\DeclareMathOperator*{\argmax}{arg\,max}

\usepackage{listings}
\lstdefinelanguage{Ini}
{
    basicstyle=\ttfamily\small,
    columns=fullflexible,
    morecomment=[s][\color{blue}\bfseries]{[}{]},
    morecomment=[l]{;},
    commentstyle=\color{gray}\ttfamily,
    morekeywords={},
    otherkeywords={=},
    keywordstyle={\color{green}\bfseries}
}
\lstdefinelanguage{BNF}
{
    basicstyle=\ttfamily\small,
    columns=fullflexible,
    morecomment=[s][\color{blue}\bfseries]{<}{>},
    morecomment=[l]{//},
    commentstyle=\color{gray}\ttfamily,
    morekeywords={},
    otherkeywords={::=},
    keywordstyle={\color{green}\bfseries}
}

\usepackage{multirow}
\usepackage{diagbox}

\usepackage{float}

\usepackage{todonotes}

\usepackage{csvsimple}
\usepackage{url}

\usepackage{graphicx}
\usepackage{caption}
\usepackage{subcaption}

\title{A Generic Framework for Hidden Markov Models on Biomedical Data}

\author{
\IEEEauthorblockN{Richard Fechner\IEEEauthorrefmark{1}\IEEEauthorrefmark{2}
, Jens Dörpinghaus\IEEEauthorrefmark{1}\IEEEauthorrefmark{3}\IEEEauthorrefmark{4}, Robert Rockenfeller\IEEEauthorrefmark{3}\IEEEauthorrefmark{5}\IEEEauthorrefmark{6}}, Jennifer Faber\IEEEauthorrefmark{4}

\IEEEauthorblockA{%
\IEEEauthorrefmark{1} Federal Institute for Vocational Education and Training (BIBB), Bonn, Germany}\\
\IEEEauthorrefmark{2}  University of Tübingen, Germany,\\
\IEEEauthorrefmark{3}  University of Koblenz, Germany,\\ 
\IEEEauthorrefmark{4} German Center for Neurodegenerative Diseases (DZNE)\\
\IEEEauthorrefmark{5}  School of Biomedical Sciences, University of Queensland, Brisbane, Australia\\
\IEEEauthorrefmark{6} School of Science, Technology and Engineering, University of the Sunshine Coast, Queensland, Australia}

\begin{document}
\maketitle      
\begin{abstract} 

\textbf{Background} 
Biomedical data are usually collections of longitudinal data assessed at certain points in time. Clinical observations assess the presences and severity of symptoms, which are the basis for description and modeling of disease progression. Deciphering potential underlying unknowns solely from the distinct observation would substantially improve the understanding of pathological cascades. Hidden Markov Models (HMMs) have been successfully applied to the processing of possibly noisy continuous signals. The aim was to improve the application HMMs to multivariate time-series of categorically distributed data. Here, we used HHMs to study prediction of the loss of free walking ability as one major clinical deterioration in the most common autosomal dominantly inherited ataxia disorder worldwide.
We used HHMs to investigate the prediction of loss of the ability to walk freely, representing a major clinical deterioration in the most common autosomal-dominant inherited ataxia disorder worldwide.

\textbf{Results} 
We present a prediction pipeline which processes data paired with a configuration file, enabling to construct, validate and query a fully parameterized HMM-based model. In particular, we provide a theoretical and practical framework for multivariate time-series inference based on HMMs  that includes constructing multiple HMMs, each to predict a particular observable variable. Our analysis is done on random data, but also on biomedical data based on Spinocerebellar ataxia type 3 disease.

\textbf{Conclusions} 
HHMs are a promising approach to study biomedical data that naturally are represented as multivariate time-series. Our implementation of a HHMs framework is publicly available and can easily be adapted for further applications.

\end{abstract}

\section{Introduction}

\begin{figure}[t]

        \centering

            \includegraphics[width=.45\textwidth]{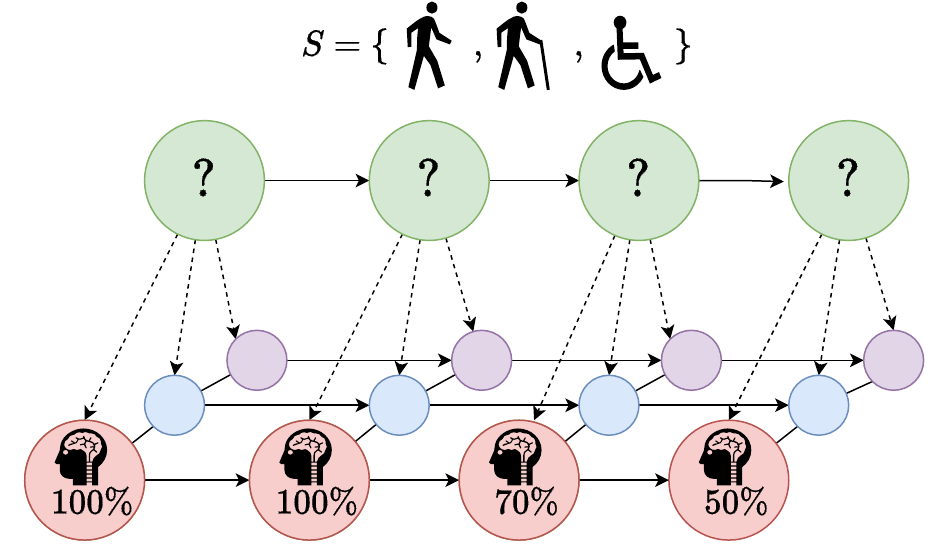}
        \quad
            \includegraphics[width=.45\textwidth]{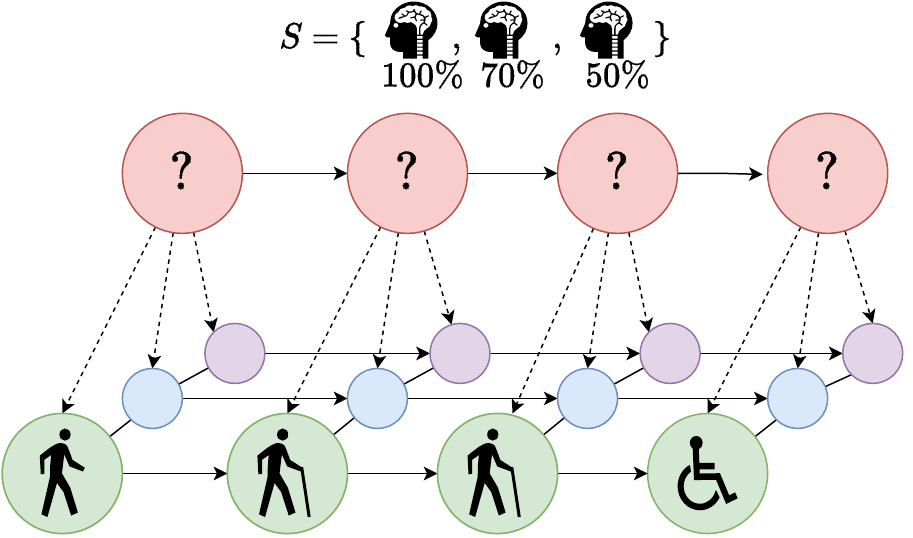}
        \caption{Conceptual depiction of the flexibility enabled by the proposed model. The predicted variable as well as the means by which to make a prediction are chosen freely by the user.}
        \label{fig:intro_1}
\end{figure}

The central idea of Machine Learning (ML) is the attempt to infer the unknown solely from observation. Every environment emits signals. In the context of health and disease, observational signs and symptoms represent such measurable signals. Here, a central question is the identification of temporal order and predictive features for deterioration. The use case for real world data we present here, are observational data from large natural history studies in the worldwide most common autosomal dominantly inherited ataxia disorder, spinocerebellar ataxia type 3, see \cite{Faber2021,ESMINfLSerum,ESMINfLPlasma,ATXN3}. Spinocerbellar ataxia 3 is a neurodegenerative disease with onset of symptoms in adult life, around the 4th decade, see \cite{SCAreview2018,SCAreview2019}. Clinical hallmarks are the progressive loss of balance, coordination deficits and slurred speech. SCA3 patients experience significant restrictions of mobility and communicative skills. Notably, SCA3 is a so-called rare disease with a prevalence of < 3 per 100,000. 
Ataxia as the most prominent symptom is measure with a scale that assesses 8 different items, like gait, stance or speech. In the observational studies included here, participants were assessed on an annual basis and the resulting scoring (itemwise as well as the overall ataxia severity sum score) are considered as signals. Measuring signals for consecutive discrete time steps present so-called multivariate time-series -- or to put it in other words -- a series of multiple observable variables over a period of time. Our aim is to infer knowledge about a state-sequence of an unknown variable from these sequences of observations. The methodological novelty presented in this article is the capability of choosing, which unknown quantity, to infer from observation. The presented model is not only able to infer knowledge about quantity $A$ by observing quantity $B$ but allows for the exploration in the opposite direction, namely reasoning about $B$ from observing $A$ (see Figure \ref{fig:intro_1}). Furthermore, our work extends this conceptual principle to a more general case, in which the model allows for maximum flexibility and generality by being able to choose any constellation of different observable variables to infer knowledge about any other observable variable. This extension is realized within the widely used probabilistic framework of Hidden Markov Models (HMMs). Our work presents two test environments, including the application of the model to real world medical data based on SCA3.

{}

HMMs have been successfully applied to the processing of possibly noisy continuous signals in the case of speech or gesture recognition \cite{baker1975dragon,nilsson2002speech, lee1999hmm}, as well as to the processing of sequences of discrete signals like text categorization \cite{frasconi2001text}. In the context of biomedical data, they have been incorporated into forecast models, improving estimates about patient mortality \cite{vairavan2012prediction}. Here, we will focus on multivariate time-series of categorically distributed data. We present a prediction pipeline which processes data paired with a configuration file, enabling to construct, validate and query a fully parameterized HMM-based model. Specifically, we will propose a rather new technique that includes constructing multiple HMMs -- one for each variable sequence of the multivariate sequence -- to predict another observable variable. We will demonstrate in a use case that the newly proposed technique can perform well when tested on real world application data. Finally, we will discuss strengths and weaknesses of the model and how the approach could be augmented to improve results.

\begin{figure}[t]
    \centering
    \includegraphics[width=0.9\textwidth]{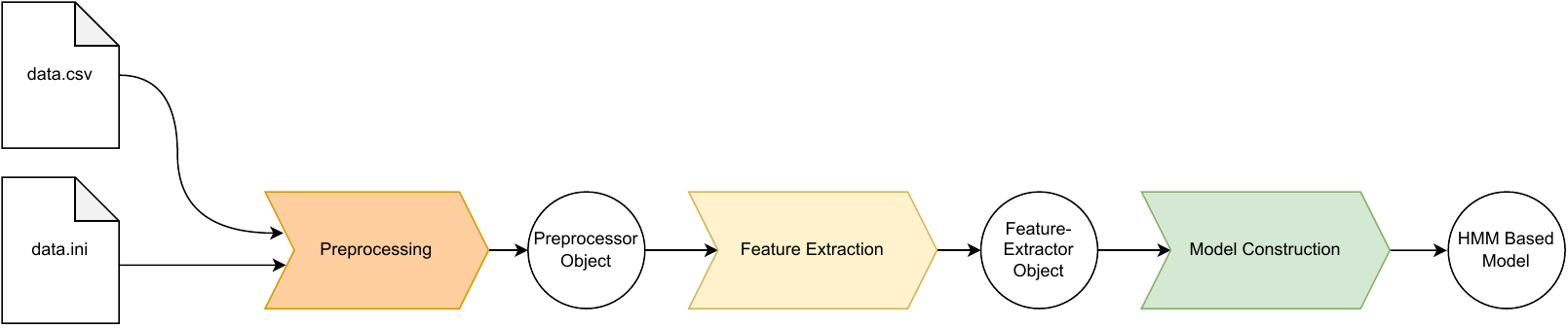}
    \caption{Concept of the presented prediction pipeline, allowing for HMM-based model construction and model query.}
    \label{fig:intro_2}
\end{figure}

The aim of this article is thus to a) provide a theoretical and practical framework for multivariate time-series inference based on HMMs and b) apply and validate the proposed prediction pipeline (see Figure \ref{fig:intro_2}) to real world data.

{}


\section{Background}\label{section:Foundations}

\subsection{Markov Chains}

A first order \textbf{Markov Chain} is a stochastic process  describing the transition between a system's states.

\begin{definition}
A Markov Chain is defined by the set of $N$ states $S = \{S_1, ..., S_N\}$ and a $N \times N$ row-stochastic transition matrix $\mathcal{A}$ where $a_{ij}$ defines the probability of transitioning from state $S_i$ to state $S_j$ in a single time step.
\end{definition}

We will be concerned with first order Markov Chains, which satisfy the so called \emph{Markov Property}.

\begin{definition}
The \textbf{Markov Property}
\begin{equation*}
    P[q_t = S_i \hspace{0.2cm}|\hspace{0.2cm} q_{t-1} = S_j] = P[q_t = S_i \hspace{0.2cm}|\hspace{0.2cm} q_{t-1} = S_j, ..., q_{t-k} = S_x]
\end{equation*}
states that the probability for state transition is solely dependent on the previous state, regardless of any other $k$ states visited before that. Here, $q_t, q_t-1$ and $q_t-k$ denotes the chain's state at time step $t$, where $S_i$, $S_j$ and $S_x$ are sample states of the underlying system. 
\end{definition}

\subsection{From Markov Chain Model to Hidden Markov Model}

In some cases, the states in our Markov Chain Model are directly observable. However, suppose that the states we would like to reason about are not directly observable and are hidden. This motivates the extension of a Markov Chain Model to a \emph{Hidden Markov Model}, in which the actual states we would like to reason about are hidden and cannot be observed. We can only observe \emph{emission-signals}, which are being emitted from the hidden states. We follow Rabiners definition of HMMs, see \cite{rabiner1989tutorial}.

\begin{definition}
A \textbf{Hidden Markov Model} $\lambda = (\mathcal{A}, \mathcal{B}, \pi)$ with the hidden states $S = \{S_1, ..., S_N\}$ and emission signals $V = \{V_1, ..., V_M\}$ is a stochastic process, defined by three model parameters
\begin{enumerate}
    \item $N \times N$ row-stochastic \textbf{state transition matrix} $\mathcal{A}$, where the element $a_{ij}$ denotes the state transition probability from state $S_i$ to state $S_j$.\\

    An alternative notation for denoting the state transition from $q_{t-1}$ to $q_t$ is given by $a_{q_{t-1}q_t}$.\\

    \item $N \times M$ row-stochastic \textbf{state emission matrix} $\mathcal{B}$ where the element $b_{ij}$ denotes the probability $P[o_t = V_j | q_t = S_i]$ of observing emission signal $o_t = V_j$ at time step $t$ under the hidden state $S_i$.\\

    An alternative notation for denoting the observation of signal $o_t$ from the hidden state $q_t$ is given by $b_{q_t}(o_t)$.\\
    \item $1 \times N$ \textbf{initial state distribution vector} $\pi$, where $\pi_i = P[q_1 = S_i]$.\\

    An alternative notation for denoting the initial state probability for hidden state $q_t$ is given by $\pi_{q_t}$.\\
\end{enumerate}
\end{definition}



Situations, where one needs to infer a sequence of hidden states from an observation sequence are very common. In the medical context for example we might be interested in inferring a sequence of diagnoses from the patients observed heartbeat and breathing frequency at certain time points.

\subsection{Three Basic Problems for HMMs}

Generally, we are concerned with inferring a sequence of hidden states from a sequence of observed states, a so-called \emph{sequence-to-sequence} prediction. The Hidden Markov Model excels at this task, although we will see that the path towards a stable prediction poses some challenges and conceals certain pitfalls one needs to avoid. Following Rabiner \cite{rabiner1989tutorial}, we see three main problems have to be solved to be able to apply HMMs in practice. We will discuss their solution within the next sections.

\begin{itemize}
\item \textbf{Problem 1}: Given an observation sequence $\mathcal{O} = o_1 o_2 ... o_T$, as well as a fully parameterized model $\lambda = (\mathcal{A}, \mathcal{B}, \pi)$, how do we calculate the probability $P[\mathcal{O}|\lambda]$ in an efficient manner?\\
\item \textbf{Problem 2}: Given an observation sequence $\mathcal{O} = o_1 o_2 ... o_T$ as well as a fully parameterized model $\lambda = (\mathcal{A}, \mathcal{B}, \pi)$, how do we find the state sequence $\mathcal{Q} = q_1 q_2 ... q_T$ best explaining the seen observation?\\
\item \textbf{Problem 3}: Given a model $\lambda$, how do we train the model, changing its parameters to maximize $P[\mathcal{O} | \lambda]$?
\end{itemize}
Solving the first problem enables us to compare different Hidden Markov Models. Given an observation sequence $\mathcal{O}$, we might prefer the model $\lambda$, which maximizes $P[\mathcal{O}|\lambda]$, the probability of observing the given sequence based on model parameters.

A solution to the second problem allows for an inference of a sequence of hidden states $\mathcal{Q}$ given an observation sequence $\mathcal{O}$. Since many different hidden state sequences might produce the given observation sequence, the problem becomes finding a $\mathcal{Q}$, which maximizes $P[\mathcal{O}|\mathcal{Q},\lambda]$, the probability of the hidden state sequence $\mathcal{Q}$ emitting the observation sequence $\mathcal{O}$.

The third problem is to approximate the in practice \emph{unknown} model parameters $\mathcal{A}, \mathcal{B}, \pi$ given an observation sequence $\mathcal{O}$. This can be interpreted as training an HMM on an observation sequence.

\subsubsection{Solution to Problem 1}

This problem is best approached naively at first, without taking computational costs into account. As we will see, the need for a smarter, less computationally intensive solution will arise along the way.

First, consider a fixed state sequence $Q = q_1...q_T$ of some states that might emit the observation sequence. The probability of this sequence arising from the model is given by the product of the probability of starting in the state $q_1$, which is given by $\pi_ {q_1}$ and the correct state transition probabilities.

\begin{equation*}
    P[Q|\lambda] = \pi_{q_1}\prod_{t = 1}^{T-1}a_{q_tq_{t+1}} = \pi_{q_1}\cdot a_{q_1q_2}...\cdot a_{q_{T-1}q_T}.
\end{equation*}
 Additionally, the probability of observing the observation sequence $\mathcal{O}$ from the state sequence $Q$ is given by the product of the single emission probabilities for the individual observations under the hidden state.

 \begin{equation*}
     P[\mathcal{O}|Q, \lambda] = \prod_{t = 1}^{T}b_{q_t}(o_t) = b_{q_1}(o_1) \cdot ... \cdot b_{q_T}(o_T)
 \end{equation*}

 Finally, the probability for observing $\mathcal{O}$ under $Q$ is $P[Q|\lambda] \cdot P[\mathcal{O}|Q, \lambda]$. Having computed this probability for one possible state sequence, all that is left is computing it again for every single possible state sequence of length $T$.

 \begin{equation*}
 P[\mathcal{O}|\lambda] = \sum_{Q \in \mathcal{Q}}P[Q|\lambda] \cdot P[\mathcal{O}|Q, \lambda]
 \end{equation*}

 Where, $\mathcal{Q}$ is the set of all possible state sequences of length $T$. Although very declarative, this solution is practically useless since the computational effort required to solve for only one observation sequence increases exponentially with the length $T$ of the sequence, as for every time step there are up to $N$ different state transitions to be made, resulting in a total of at worst $N^{T}$ different candidates for $Q$, which are to be evaluated. This is unfeasible for even a small number of hidden states $N$ and a short sequence length.

 To overcome this hurdle, we make use of dynamic programming and temporarily store intermediate results to bootstrap and extend for a new, more optimal iteration of an intermediate result until we have reached the desired solution.

\begin{definition}
The \textbf{forward variable} $\alpha_t(i) = P[o_{1}...o_t | q_t = S_i, \lambda]$ is defined as the probability of observing the partial observation sequence $o_1...o_t$ given State $S_i$ at time step $t$ as well as a fully parameterized model $\lambda$.
\end{definition}

\begin{figure}[t]
    \centering
    \includegraphics[width=0.45\textwidth]{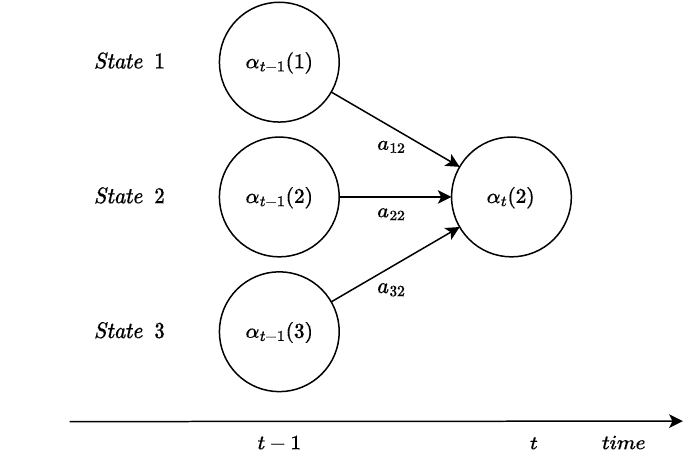}
    \caption{Usage of dynamic programming in the forward algorithm.}
    \label{fig:lattice_forward}
\end{figure}

 We will use dynamic programming to compute the forward variable $\alpha_t(i)$ from our solutions for $\alpha_{t-1}(j)$, where $1 \leq j \leq N$ (see Figure \ref{fig:lattice_forward}). The full solution is as follows:\\

 {}

 \begin{enumerate}
     \item Initialization: $          \alpha_1(i) = \pi_i \cdot b_i(o_1) \hspace{1cm} 1 \leq i \leq N $
     \item Induction:
        \begin{align*}
            \alpha_t(j) = \left(\sum_{j = 1}^N \alpha_{t-1}(j)a_{ij}\right) \cdot b_j(o_{t-1})
                            & \hspace{1cm} 1 \leq j \leq N\\
                            & \hspace{1cm} 1 < t \leq T
         \end{align*}
     \item Termination: $P[\mathcal{O} | \lambda] = \sum_{i = 1}^N \alpha_T(i)$
 \end{enumerate}{}

 \begin{figure}[t]
    \centering
    \includegraphics[width=0.9\textwidth]{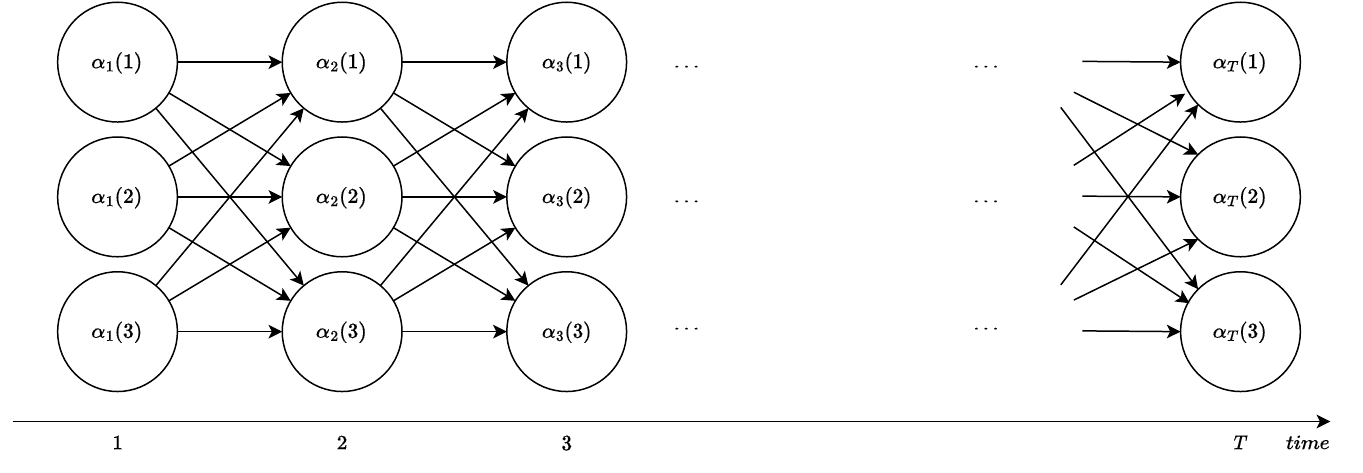}
    \caption{Visual representation of the full lattice structure used to compute $\alpha_T(i)$, with $1 \leq i \leq N$.}
    \label{fig:full_lattice}
\end{figure}{}

In the first step, we initialize the storage for the intermediate results, before continuously calculating the next intermediate results (see Figure \ref{fig:full_lattice}). In the end, we sum over $\alpha_T(i)$ to obtain our desired result. This method of bootstrapping and reusing old results dramatically reduces the number of required operations down to $N^2T$ (from the previous $N^T$) operations \cite{rabiner1989tutorial}. It should be noted that computing $\alpha_t(i)$ for every $t$ and $i$ for a given observation sequence $\mathcal{O}$ yields the so-called posterior distribution for the hidden states given the observation sequence. This will be important later on when we will discuss the prediction capabilities.

\subsubsection{Solution to Problem 2}

One possible solution to the question ``\textit{What's the most likely state-sequence for observation $\mathcal{O}$?}'' is to find the most probable state $S_i$ for each time step $t$. To tackle this problem we will need further definitions, first of all, let us define the so-called ``backward-variable'', which in its definition is quite similar to the forward variable.

\begin{definition}
The \textbf{backward variable} $\beta_t(i) = P[o_{t+1}...o_T | q_t = S_i, \lambda]$ is defined as the probability of observing the partial observation sequence $o_{t+1}...o_T$ given State $S_i$ at time step $t$ as well as a fully parameterized model $\lambda$.
\end{definition}

\begin{figure}[t]
    \centering
    \includegraphics[width=0.45\textwidth]{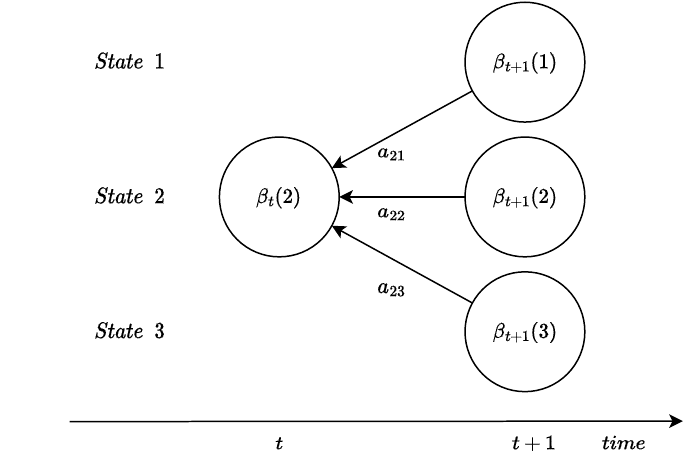}
    \caption{Usage of dynamic programming for the calculation of the backward variables.}
    \label{fig:lattice_backward}
\end{figure}{}

To calculate the backward variable, we use the same idea of dynamic programming, reusing former results (see Figure \ref{fig:lattice_backward}). Although this time, we travel ``backwards'' through the observation sequence, thus starting at the last observation $o_T$.

\begin{enumerate}
     \item Initialization: $         \beta_T(i) = 1 \hspace{1cm} 1 \leq i \leq N$
     \item Induction:
        \begin{align*}
            \beta_t(i) = \sum_{j = 1}^N \beta_{t+1}(j)a_{ij}b_j(o_{t+1})
                            & \hspace{1cm} 1 \leq i \leq N\\
                            & \hspace{1cm} 1 \leq t < T
         \end{align*}
 \end{enumerate}{}

Having both, the forward and the backward variable at our disposal, we can continue defining a helper variable $\gamma_t(i)$.

\begin{definition}
The \textbf{helper variable} $\gamma_t(i) = P[q_t = S_i | \mathcal{O}, \lambda]$ denotes the probability of being in the hidden state $S_i$ at time step $t$ given the full observation sequence $\mathcal{O}$ as well as a fully parameterized model $\lambda$.
\end{definition}

We can express the variable $\gamma_t(i)$ in terms of the forward and backward variables (see Figure \ref{fig:lattice_forward_backward}): $    \gamma_t(i) = \frac{\alpha_t(i)\beta_t(i)}{P[\mathcal{O}|\lambda]} =     \frac{\alpha_t(i)\beta_t(i)}{\sum_{i=1}^N \alpha_t(i)\beta_t(i)}
$.
In this equation $P[\mathcal{O}|\lambda]$ is a normalization factor. Thus, $\sum_{i=1}^N \gamma_t(i) = 1$. To continue with our interpretation of optimality for a state sequence, we can solve for the individually most likely states for each time step:
    $q_t = \argmax_{1 \leq i \leq N}(\gamma_t(i)),\;  1 \leq t \leq T$.
Unfortunately, there is a flaw in this solution. Although we have found the individually most likely states, we have no guarantee for soundness of the found sequence of states. It might just be that our sequence contains an illegal state transition. This error is resolved by the \textit{Viterbi Algorithm} \cite{viterbi1967error}:

\begin{figure}
    \centering
    \includegraphics[width=0.9\textwidth]{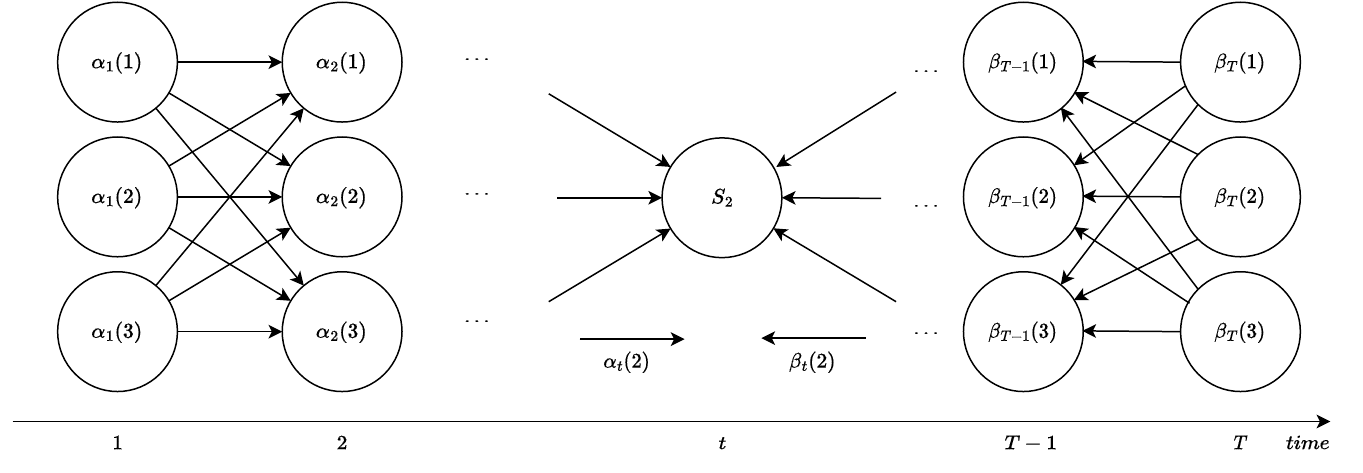}
    \caption{Usage of lattice structure in the forward-backward algorithm.}
    \label{fig:lattice_forward_backward}
\end{figure}{}

\begin{definition}
The \textbf{Viterbi Algorithm} finds the most likely state sequence $Q = \{q_1,q_2,...,q_T\}$ to have emitted a given observation sequence $\mathcal{O} = \{o_1,o_2, ...,o_t\}$, whilst respecting the state transition constraints given by the model $\lambda$.
\end{definition}

\begin{definition}
$\delta_t(i)  = \max\limits_{q_1,...,q_{t-1}} P[q_1,...,q_t = i, o_1,...,o_t | \lambda]$ is the probability for the most probable state sequence along a single path $q_1,...,q_t$ with observation sequence $o_1,...,o_t$ up until time step $t$ that ends in the hidden state $S_i$.
\end{definition}{}

Again, we can compute this variable inductively by $\delta_{t}(j) = (\max_{1 \leq i \leq N}\delta_{t-1}(i)a_{ij})\cdot b_j(o_t)$.
Since our goal is to find the optimal state sequence, we need a way to keep track of which prior state maximized $\delta_t(i)$ at each time step $t$ for each state $i$. This is done with a storage array $\psi_t(i)$. This lets us define the Viterbi Algorithm in four steps.

\begin{enumerate}
    \item Initialization:
    \begin{align*}
        & \delta_1(i) = \pi_ib_i(o_1), \hspace{1cm} 1 \leq i \leq N\\
        & \psi_1(i) = 0
    \end{align*}{}

    \item Recursion:
    \begin{align*}
        & \delta_t(j) = \max_{1 \leq i \leq N}(\delta_{t-1}(i)a_{ij})b_j(o_t) &1 \leq j \leq N\\
        &&2 \leq t \leq T\\
        & \psi_t(j) = \argmax_{1 \leq i \leq N}(\delta_{t-1}(i)a_{ij}) &1 \leq j \leq N\\
        &&2 \leq t \leq T
    \end{align*}{}

    \item Termination:
    \begin{align*}
        & P^* = \max_{1 \leq i \leq N}(\delta_T(i))\\
        & q_T^* = \argmax_{1 \leq i \leq N}(\delta_T(i))
    \end{align*}{}

    \item Backtracking:
    $q_t^* = \psi_{t+1}(q_{t+1}^*)$, \hspace{0.2cm} $t = T-1,T-2,...,1$

\end{enumerate}{}


\subsubsection{Solution to Problem 3}

Problem 3, which is concerned with training an HMM in a way such that the model is more likely to explain a given observation sequence, is the most difficult one out of the three presented problems. Starting from a rough estimate of the model parameters $\mathcal{A},\mathcal{B}, \pi$, we can iteratively improve the model with the Baum-Welch Algorithm, which can be interpreted as an application of the Expectation-Maximization-Algorithm to HMMs \cite{dempster1977maximum}, see also \cite{rabiner1989tutorial} for more details.

\subsection{Machine Learning Metrics}

Metrics are an important tool in ML to quantify the success of a system for a given dataset. They are a measure of quality for the learned model. Additionally, metrics allow the comparison of different models with different architectures. The metric used is dependent on the type of prediction that is to be made and since the following classification problem presented is of a multi-class nature. We will motivate and explain one suitable metric for these classification problems -- the multi-class $F_1$-Score.



\begin{definition}
We call $    F_{\beta} = (1 + \beta^2) \cdot \frac{precision \cdot recall}{(\beta^2 \cdot precision) + recall}$
the $F_{\beta}$-Score. For $\beta = 1$ we obtain the $F_1$-Score.
\end{definition}

The $F_1$-Score is a common metric used to measure the quality of binary classifiers. It represents the harmonic mean between precision and recall, two values we can compute with the help of $C$, our confusion matrix. $precision = \frac{true\hspace{0.2cm} positives}{true\hspace{0.2cm} positives + false\hspace{0.2cm} positives}$, and $recall = \frac{true\hspace{0.2cm} positives}{true\hspace{0.2cm} positives + false\hspace{0.2cm} negatives}$.
We can extend the $F_1$-Score to the multi-class domain, by computing a $K \times K$ confusion matrix, where $K$ is the number of classes our model can predict. 
%
%
Thus, we can compute the $F_1$-Scores for each class. The final $F_1$-Score is either an average or a weighted sum of the respective $F_1$-Scores \cite{grandini2020metrics}.\\


\section{Method}
\label{section:Method}

In this  section, we will introduce the notation that will be used, and give an in-depth model overview.

\subsection{Notation}

\begin{definition}
A \textbf{marker $M$} is a unique identifier or class-name for a certain set of discrete states $S = \{S_1,...,S_N\}$. We say that $\mathcal{M}$ is the set of all markers.
\end{definition}

Markers are the abstract handle for the observable units in our environment. For example, whilst observing a medical patient, the set of markers $\mathcal{M}$ could for example include ``ability to walk'', ``blood pressure'', ``breathing frequency''. The set of concrete states of the marker ``blood pressure'' could be $S = \{ low, medium, high\}$. $\mathcal{M}$ is partitioned into a set of layers $\mathcal{L}$.

\begin{definition}
A \textbf{(prediction-) layer} $L$ is an element of $\mathcal{L}$, partition of $\mathcal{M}$. A mapping $\mathfrak{L}_{M\rightarrow{L}} : \mathcal{M} \rightarrow{\mathcal{L}}$ maps a given marker $M$ to its corresponding layer $L$.
\end{definition}

Using this definition, we can partition the set of markers into layers containing related markers. For example, we could group the markers ``ability to walk'' and ``number of push-ups'' into a layer ``mobility''. Next up, we should define the importance of these layers and their markers for our expected prediction. This is done by defining the following weight functions.

\begin{definition}
Let a \textbf{layer-weight function} $\omega_{\mathcal{L}} : \mathcal{L} \rightarrow{\mathbb{R}_{\geq 0}}$ be a mapping from a layer $L$ to a positive weight. The function $\omega_{\mathcal{L}}$ must satisfy $\sum_{L \in \mathcal{L}} \omega_{\mathcal{L}}(L) = 1$.
\end{definition}

\begin{definition}
Let a \textbf{marker-weight function} $\omega_L : L \rightarrow{\mathbb{R}_{\geq 0}}$ be a mapping from a marker $M$ of layer $L$ to a positive weight. The function $\omega_M$ must satisfy $\sum_{M \in L} \omega_L(M) = 1$.
\end{definition}

\begin{definition}
A \textbf{Trail $T=o_1...o_t$} is an observation sequence of consecutive states of length $t$, which belong to a marker $M$. The weight of a trail is defined as $\omega_T(T) = \omega_L(M) \cdot \omega_{\mathcal{L}}(L_{M\rightarrow{L}}(M))$.
\end{definition}

\begin{definition}
An \textbf{Observation} $\mathcal{O} = \{T_1, ..., T_k\}$ is a set of Trails of same length $t$. The mapping $\mathfrak{M}_{T \rightarrow{M}} : \mathcal{O} \rightarrow{\mathcal{M}}$ maps a Trail $T$ to its corresponding marker $M$.
\end{definition}

\begin{definition}
A \textbf{Query} $\mathcal{Q} = (M_{\mathcal{H}}, \mathcal{L}, \mathcal{O}, \{\omega_{L_1}, ..., \omega_{L_k} \}, \omega_{\mathcal{L}})$ consists of a hidden marker $M_{\mathcal{H}}$ as well as a partition of $\mathcal{M}$, namely $\mathcal{L}$. A Query possesses an observation $\mathcal{O}$, consisting of possibly many trails $T$. The weights of the markers and layers are defined by the marker-weight functions $\omega_{L_i}$ (one for each layer) and the layer-weight function $\omega_\mathcal{L}$.
\end{definition}

Thus, a query $\mathcal{Q}$ is a well-defined description of a multivariate sequence alongside mixture components, namely the weights of the markers and layers. The mixture components are the weights, by which the individual result of each HMM is weighted. We would like to continue to define a formalism, which maps augmented versions of these queries to a prediction result. Thereby, the form of the prediction is dependent on the augmented query. Plainly said, our model can predict time series of discrete states, as well as time series of distributions over states. At first, we will keep the prediction result -- in any case, some sort of time series -- very abstract.

\begin{definition}
A \textbf{trail-evaluation} $   \phi(T, M_{\mathcal{H}}) = \omega_T(T) \cdot P(T, \lambda(M(T), M_{\mathcal{H}}))$ is a function that maps a Trail, a Query and possibly various arguments to a specifically requested weighted prediction result $P(T, \lambda(M(T), M_{\mathcal{H}}))$.
\end{definition}
Here, $\lambda(M(T), M_{\mathcal{H}})$ is a function that returns a fully parameterized HMM whose hidden states are the states of the hidden marker $M_{\mathcal{H}}$ and the observable states are the states of the trail marker $M(T)$. Subsequently, the resulting HMM is used to reason about the given trail $T$. Further details about the nature of the prediction are given in Section \ref{section:model_validation_and_query}.
\begin{definition}
An \textbf{observation-evaluation} $   \Phi(\mathcal{Q}) = \sum_{T \in \mathcal{Q}}\phi(T, M_{\mathcal{H}}),\hspace{0.5cm}M_{\mathcal{H}} \in \mathcal{Q}$ is a function that sums up the weighted prediction results to construct the final prediction.
\end{definition}

Hence, an observation-evaluation of a query is the weighted sum of the evaluations of the trails - the constituents of the observation.

\subsection{Model Overview}
\label{section:model_overview}

Observing an environment in the real world allows for the observation of the states of multiple markers $M_i$ at each time step. Certain semantically related markers are grouped into layers. We would like to reason about a hidden state in this environment based upon our observations by constructing a query $\mathcal{Q}$. The idea is to construct an HMM for every single marker and try to infer knowledge about the hidden states from given observations, according to the weight of the marker $\omega_L(M)$ and the weight of its respective layer $\omega_{\mathcal{L}}(L(M))$. The resulting model can briefly be described as a mixture of Hidden Markov Models -- one HMM for each marker. It is important to understand that our model is only able to offer prediction capabilities that a simple one-observation HMM could offer as well.


In this work, we assume the hidden marker to be one of the visible observation markers, allowing for the ability to generically switch the hidden marker and prediction-layer to allow for maximum flexibility. Hence, we will be able to extract the parameters of the model, namely $\mathcal{A}, \mathcal{B}$, and $\pi$ from the observation sequences directly, instead of having to train the model based on said sequences. This is in stark contrast to the usual usage of HMMs, where these parameters are not given and have to be approximated (learned) from observations.

\subsection{Pipeline Description}
\label{section:pipeline_description}

\begin{figure}[t]
    \centering
    \includegraphics[width=.8\textwidth]{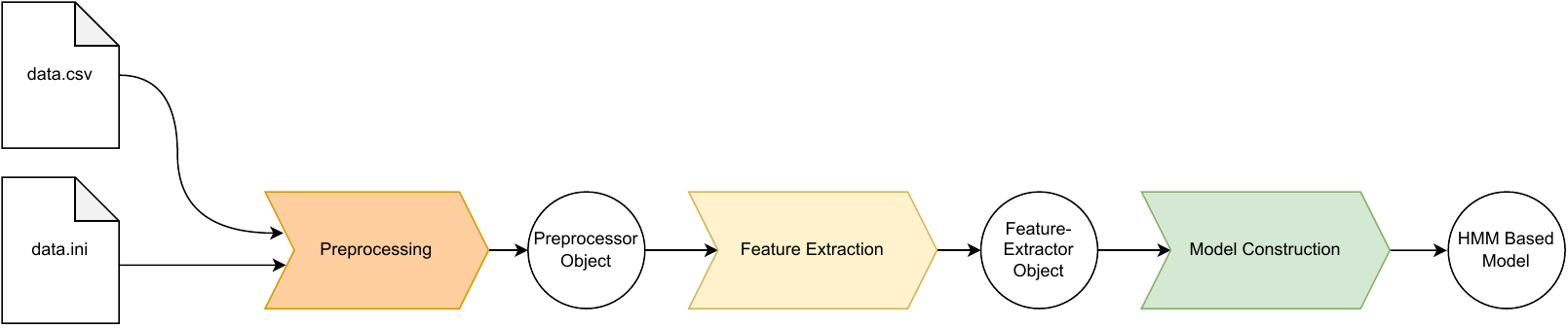}
    \caption{Flow-diagram of the prediction pipeline. The data is converted to a fully functioning model.}
    \label{fig:pipeline}
\end{figure}

The prediction pipeline was written in Python, making use of the \texttt{hmmlearn} library\footnote{The pipeline and documentation is available at \url{https://github.com/rfechner/generic-hmm}.}. The pipeline comprises a pre-processing step, a feature extraction step, and a prediction step (see Figure \ref{fig:pipeline}). To obtain a trained model, the user must input their training data in the form of a csv-file as well as an ini-config file, describing the layer structure and provide sufficient information about the weights of the model.

{}

\subsubsection{User Input}

The data consisting of the different observations provided by the csv-file, as well as the configuration of the model provided by the ini-file, must be supplied by the user.
The abstract syntax outlines the correct way of specifying an ini-file required to construct a model, whilst making no assumption about the form of the data, thus it uses the Backus Naur Form (BNF, see \cite{knuth1964backus}). Each marker in the data must be specified as a section inside the file. Additionally, information about the datatype, related layer, and layer-specific weight of the marker must be supplied as a key-value pair under the corresponding markers section. Optional information, like the relationship to other markers, can be added.

It should be noted that, although the definition of a marker calls for the existence of a layer-specific weight, the program is robust against missing or faulty weights and will re-balance the given weights to satisfy stochastic constraints.

\subsubsection{Pre-processing}

\begin{figure}[t]
    \centering
    \includegraphics[width=.8\textwidth]{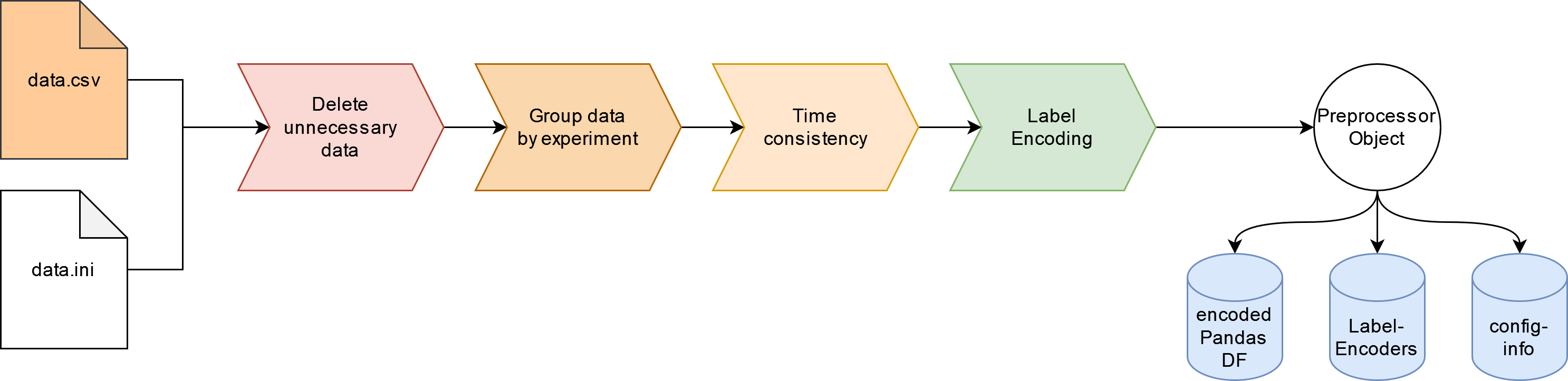}
    \caption{Flow-diagram of the pre-processing pipeline.}
    \label{fig:prepPipeline}
\end{figure}

Pre-processing is a modular stage inside the pipeline, which itself is a small pipeline (see Figure \ref{fig:prepPipeline}). The transformation of the data supplied includes the analysis of the ini-file, deletion of any unnecessary data, the grouping of the data according to the metainformation extracted from the ini-file, enforcing measurement interval consistency if necessary, and finally encoding the data into a less memory intensive format.

{}





\subsubsection{Feature Extraction}

\begin{figure}[t]
    \centering
    \includegraphics[width=.8\textwidth]{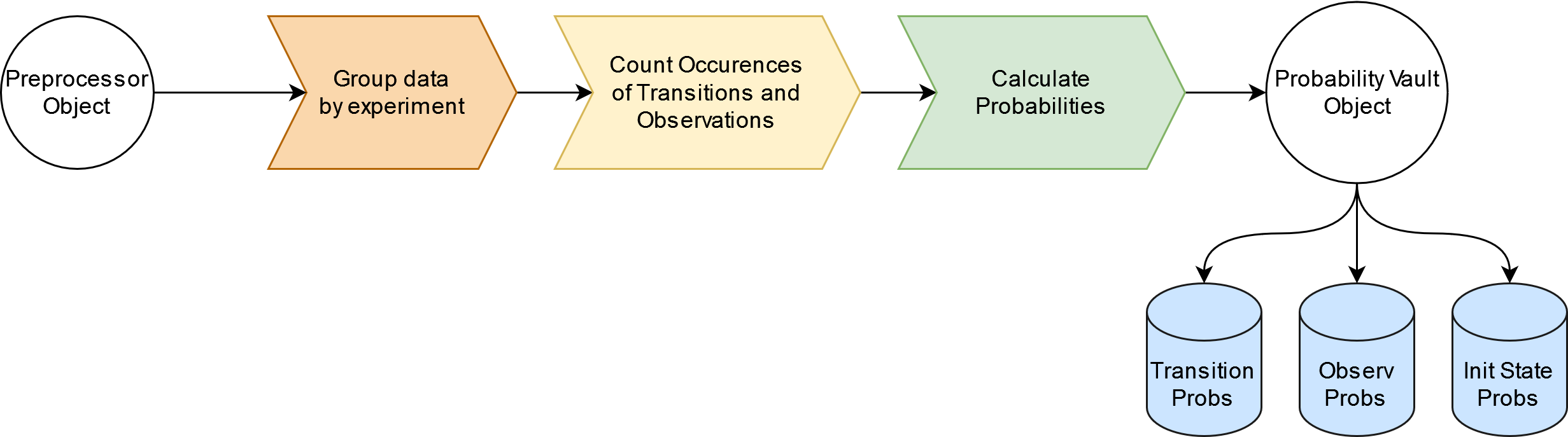}
    \caption{A flow-diagram of the Feature Extraction Pipeline.}
    \label{fig:FEpipeline}
\end{figure}

The feature extraction builds upon the previous step in the pipeline, the pre-processing (see Figure \ref{fig:FEpipeline}). Just as the prior component of the pipeline, the feature extraction component is fully modularized, implementing the necessary interface used to provide the required functionality to the next part in the pipeline. Inside the feature extraction stage, the \emph{state transition}-, \emph{signal emission}- and \emph{initial state}-probabilities are extracted from the encoded data supplied by the pre-processing stage. This is an important step since we can use the extracted probabilities later on to construct HMMs with a strong initial guess for $\mathcal{A}, \mathcal{B}$ and $\pi$.

{}

\subsubsection{Model Validation and Query}
\label{section:model_validation_and_query}

Finally, in the last step of the pipeline, the HMM-based model is constructed and queried. Building on top of the previously constructed feature extraction, the actual construction of the different HMMs is very convenient. It includes the interpolation of the results given by the individual HMMs according to the weights specified inside the configuration. To give a measure of success, the model is able to compute the multi-class $F_1$-Score for a given validation dataset. 






As already stated, it is important that the HMM-based model can only offer predictions that a simple single observation HMM could offer as well. These predictions include the following:

\begin{enumerate}
\label{section:prediction_capabilities}
    \item \textbf{Posteriors for each hidden state for observation $\mathcal{O}$}\\
    Given a Query $\mathcal{Q}$, compute the posterior distribution for each hidden state of $M_{\mathcal{H}}$ for every time step $t$ given the trails $T_i$. The result will be a weighted sum (according to the weights defined in $\mathcal{Q}$) of the individually computed posteriors.

    \item \textbf{Distribution over hidden states following $\mathcal{O}$}\\
    Given a Query $\mathcal{Q}$, predict the distribution over the hidden states of $M_{\mathcal{H}}$ for possibly many time steps $\hat{t}$ following the observation. This yields an approximation to a stationary distribution of the state transition matrix for $M_{\mathcal{H}}$. The kind of stationary distribution is dependent on the initial state distribution given by the observation sequence $\mathcal{O}$.

    \item \textbf{Optimal state sequence}\\
    Given a Query $\mathcal{Q}$, compute the optimal state sequence of $M_{\mathcal{H}}$ ``best explaining'' the trails $T_i$ using the Viterbi Algorithm.
\end{enumerate}

\subsubsection{Controller}

To wrap all of these components up, and use the whole pipeline, as well as enable plotting of the results, a wrapper object called \texttt{Controller} provides a user-friendly interface.

\subsection{Data}

\subsubsection{Data Generation}

The random dataset mimics the actual biomedical dataset in which we expect to see many markers with degenerative states -- markers, whose states continuously progress over time, going from an initial good state into a worse state. Four different degenerative markers were selected, namely

\begin{enumerate}
    \item $M_{finemotor}$, a marker whose state indicates the subject's ability to solve tasks using their hands.

    \item $M_{mobility}$, a marker whose state indicates the subject's state of mobility, e.g. still being able to walk freely without the need of walking aids or the need to use a wheelchair. 

    \item $M_{neuropsych}$, a marker whose state indicates the subject's ability to solve mental tasks.

    \item $M_{diagnosis}$, a marker whose state indicates the diagnosis given by an expert for a subject at a certain time step.
\end{enumerate}

Every marker has the same set of degenerative states, namely $S = \{$good, med-good, med, med-bad, bad, severe $\}$ indicating the current state under the given marker. To construct the data, a state transition matrix $\mathcal{A}$ as well as an initial state distribution vector $\pi$ were constructed for the single marker $M_{diagnosis}$. Additionally, emission signal matrices ($\mathcal{B}_{motoric}$, $\mathcal{B}_{mobility}$, $\mathcal{B}_{neuro}$) were constructed for the other markers $M_{motoric}$, $M_{mobility}$, $M_{neuro}$. Using these parameters, three different HMMs were constructed. Using the initial state probability distribution, for each observation sequence, an initial state for the marker $M_{diagnosis}$ was chosen. Thereafter, the following states for $M_{diagnosis}$ were sampled from the state transition probability matrix $\mathcal{A}$. Likewise, for each time step of each observation, the emission signal for each observation marker $M_i$ was sampled from the distribution located in the respective emission signal matrix $\mathcal{B}_i$. A dataset of $N = 300$ observation sequences of variable sequence-length and variable time intervals in between measurements was synthesized. To finalize the data, a configuration ini-file was constructed. For simplicity, each marker was assigned to a distinct layer, resulting in four layers -- one for each marker.

\subsubsection{Background and Description of the biomedical dataset}

Here, we will discuss a dataset obtained by a longitudinal observational study of subjects suffering from spinocerebellar ataxia type 3 (SCA3), the most common autosomal dominantly inherited neurodegenerative ataxia disorder (ESMI, European spinocerebellar ataxia type 3/Machado-Joseph disease initiative \cite{Faber2021, ESMINfLSerum,ESMINfLPlasma,ATXN3}). Symptoms include progressive loss of balance, coordination deficits and slurred speech. SCA3 patients experience significant restrictions of mobility and communicative skills. Preventive interventions that aim to silence the disease gene offer a promising treatment option. The first clinical gene silencing trial\footnote{ClinicalTrials.gov, Identifier: NCT05160558} has recently started. Consequently, there is an urgent need to predict the deterioration to a more severe disease stage to prioritize and treat patients suffering from a greater risk with less uncertainty.

The supplied data contains results from various scales that were assessed in all participants on an almost annual basis. Usually, those scales reflect increased impairment with higher scoring, while 0 refers to a normal condition and the absence of signs or symptoms like in the healthy population. In our analyses we included the following 4 assessments that measure special ends of abilities or measured observations of subjects, where a rating of 0 indicates a healthy condition: 1) disease staging (ranging from 0=normal over 1=ataxic, but able to walk freely, 2=need to use walking aids, 3=requirement to use wheelchair) \cite{diseasestages}, 2) Scale for the assessment and rating of ataxia (SARA), measuring the ataxia severity rated in 8 different items (namely gait, stance, sitting, speech, finger chase, nose-finger test, fast alternating hand movements and heel-shin slide) resulting in a sum score (raging from 0 to 40) \cite{SARA}; 3) Inventory of non-ataxia signs (INAS), assessing neurological symptoms other than ataxia \cite{INAS} and 4) a scale assessing the activities of daily life (ADL) \cite{ADL}. Notably, the  ``ADL'' (Activities of Daily Life), ``INAS'' (Intentional Non-Adherence Scale) and ``SARA'' (Scale for the Assessment and Rating of Ataxia) scales were considered in the first part of the experiment. These specific scales were chosen after consulting with a domain expert. With the additional information about the -- by domain experts already inferred -- disease state for each multivariate observation in our data, we return to the familiar setup presented in the prior section, where we have multiple trails paired with a given diagnosis or inferred latent state. $M_{diagnosis}$ in this context represents the disease stage (normal, ataxic, walking aids, wheelchair). A ``diagnosis'' in its native meaning usually describing a particular disease does not make sense in the context of a hereditary disease that is ``diagnosed'' simply by a genetic test. Thus, instead of the 3 observation markers like in the prior example, we here get about 109 different markers, which we can observe at every time step. The average length of the given time-series was calculated to be about 2.3 time steps.

\section{Results}
\label{section:Evaluation}

The model was evaluated on two different datasets:  a random dataset and a real-world biomedical dataset containing observational data of a longitudinal natural history study in spinocerebellar ataxia type 3 (SCA3). Due to data-protection reasons, we will present an in-depth discussion of results and evaluation for the first dataset.

\subsection{Evaluation 1: Random Data}

The dataset was evaluated with a 10-fold cross validation. The layers of query $\mathcal{Q}$ were set to $\mathcal{L} = \{$ $L_1$ = mobility, $L_2$ = motoric, $L_3$ = neuro $\}$. In this special case, every marker has an equal influence on the model prediction, since every layer consists of only one marker and the layer weights are equally distributed across all layers. Alternatively, the layer weights $w(L_i)$ could be adjusted, or multiple markers could be grouped in one layer, allowing for further differentiation by adjusting the marker weights $w(M_i)$ inside one layer. The hidden marker $M_{\mathcal{H}}$ was set to the marker $M_{diagnosis}$.\\

\begin{figure}[t]
    \centering
        \centering
            \includegraphics[width=.75\textwidth]{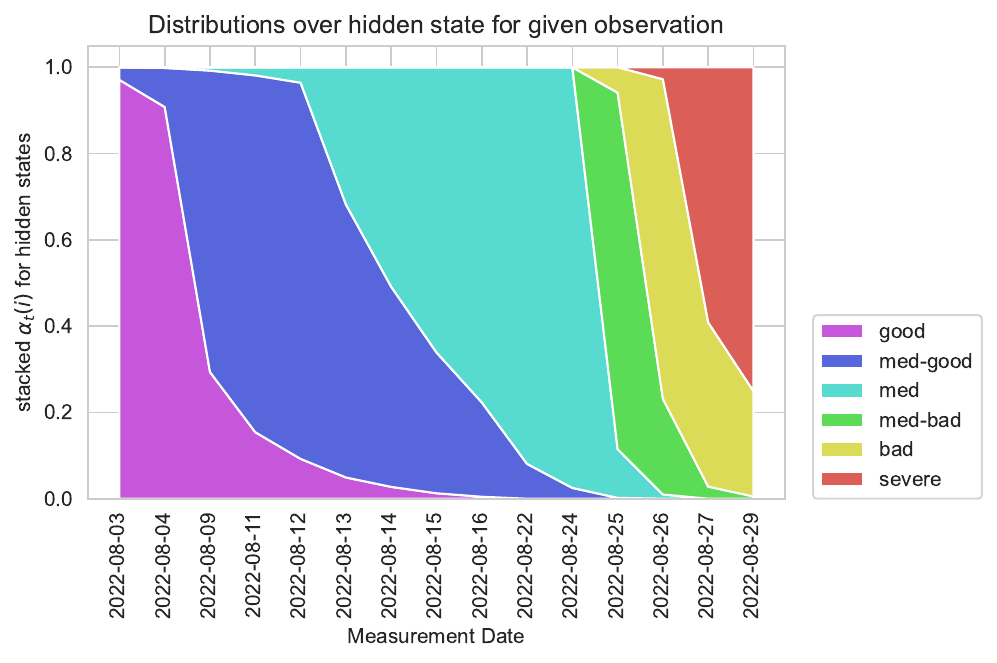}

            \includegraphics[width=.75\textwidth]{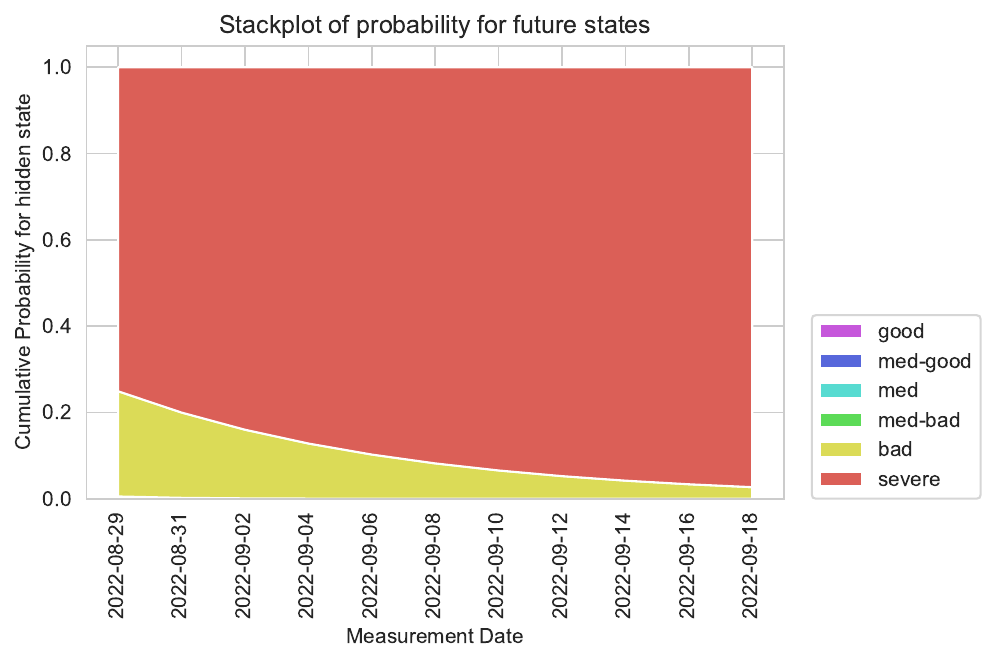}
    \caption{The distribution over the posteriors for the given hidden states (top) and the extrapolated distribution (bottom) under $M_{\mathcal{H}} = M_{diagnosis}$ changes over time.}
    \label{fig:first_dataset}
\end{figure}

As already stated above, the model was evaluated with a 10-fold cross validation, which yielded an average \textbf{$\boldsymbol{F_1}$-Score of 0.813 with a standard deviation of 0.017}. To showcase the prediction capabilities of the model further, a never seen before observation was generated, and the model was queried for all three possible prediction types (see Section \ref{section:model_validation_and_query}). These include the prediction of the distribution over the posteriors and an extrapolation for the posterior distributions given an observation sequence (see Figure \ref{fig:first_dataset}).

{}

\subsection{Evaluation 2: Biomedical Data}

\subsubsection{First Experiment}

The 109 markers were partitioned into three layers corresponding to the scale each marker belongs to; $L_{ADL}, L_{INAS}$ and $L_{SARA}$. The weight function $\omega_{L}$ assigns each prediction-layer the weight $\frac{1}{3}$. Internally, the weight function responsible for assigning a weight to every marker inside a given layer assigns equal weight to every marker. The hidden marker $M_{\mathcal{H}}$ was chosen to be the diagnosis marker. This proved to be a naive approach, as the bad performance of the model with an $F_1$-Score of less than 0.5 in a 10-fold cross-validation of the whole dataset reassured.

\subsubsection{Second Experiment}
\label{section:summary_of_results}

Since first experiments yielded an unacceptable performance, the decision was made to reduce the number of Trails, by only considering the cumulative score for each of the three scales (ADL, INAS and SARA), drastically reducing the number of Trails down to only three. Within this reduced setting, the model performed relatively well, with a mean $F_1$-Score of 0.75 and a standard deviation of 0.04 as the result of a 10-fold cross validation on the whole dataset. Additionally, further constellations of prediction-layers and hidden markers were tested (see Table \ref{tab:results}). The predictions with $M_{\mathcal{H}}=$``diagnosis'' generally produced acceptable results, whereas the other predictions yielded rather unsatisfying prediction results.

\begin{table}[]
    \centering
    \begin{tabular}{|l||*{4}{c|}}
        \hline
        \multicolumn{5}{|c|}{$F_1$-Scores for 10-fold cross validation ($\mu \pm \sigma$)} \\
        \hline
        \backslashbox[40mm]{layers}{$M_{\mathcal{H}}$} &diagnosis&ADL score&INAS score&SARA score\\
        \hline
        $L_{ADL}, L_{INAS}, L_{SARA}$& 0.75 $\pm$ 0.04 & 0.22 $\pm$ 0.06 & 0.13 $\pm$ 0.04 & 0.19 $\pm$ 0.06\\
        $L_{SARA}$&   0.75 $\pm$ 0.03  & 0.23 $\pm$ 0.03   & 0.16 $\pm$ 0.04 & -\\
        $L_{INAS}$& 0.6 $\pm$ 0.06  & 0.11 $\pm$ 0.04   & - & 0.13 $\pm$ 0.04\\
        $L_{ADL}$& 0.72 $\pm$ 0.08  & -   & 0.19 $\pm$ 0.05 & 0.24 $\pm$ 0.04\\
        \hline
    \end{tabular}
    \caption{Table cells display the mean and standard deviation of the $F_1$-Scores received from a 10-fold cross validation. Different constellations of prediction-layers and hidden marker were used. Overall on the level of single layers LSARA allows the highest prediction of disease staging $M_{diagnosis}$ (normal, ataxic, walking aid, wheelchair), followed by LADL, representing the activities of daily living, while neurological symptoms other than ataxia (LINAS) do not allow any meaningful prediction. }
    \label{tab:results}
\end{table}{}

\section{Discussion}

Generally, it was expected that the model would perform relatively well given the prediction right prediction-layers and enough training data. A substantial difficulty with the real-world data we had access to, was that a majority (approx. $\frac{2}{3}$) of the given training sequences was of length 1 or 2, thus the dataset lacked sufficient samples for longer time-series. Naturally, this also effected the validation dataset. This leads to an important question: Which clinical trials or patient studies produce data suitable for an HMM-based prediction? It was shown in Section \ref{section:summary_of_results} that we can reach reasonable prediction scores using an HMM approach -- at least for the right constellations of prediction-layers and hidden marker.

The reasonable results produced, given the prediction-layers $L_{ADL}$, $L_{INAS}$, $L_{SARA}$ in constellation with the hidden marker $M_{\mathcal{H}}$ set to the marker ``diagnosis'' can be explained by the strong correlation between the individual markers inside the prediction-layers (which are simply the scores for the corresponding scale) and the diagnosis given by a medical professional. Specifically, the prediction-layer $L_{SARA}$ seems to be the best suitable for producing predictions for the hidden marker ``diagnosis''. This independently received finding correlates with the findings of domain experts, i.e. the biomedical experts.

We also need to discuss the limitations of our approach. 
First, as discussed above, the time-series as base for prediction, as well as the distribution for the initial hidden state have to be considered. As already described, about $\frac{2}{3}$ of the training data contains time-series only of length 1 or 2. This greatly influences the prediction capabilities for longer time-series. The dataset does not have enough training samples for longer time-series, thus the performance for predicting longer sequences could be better.
Additionally, since we extract the probabilities from the training data, the model has a strong bias for the probability of the initial state. Without knowing anything about the validation sample, a certain initial state is far more probable than another. This fact, paired with the appearance of short (sometimes single time step) time-series, is most definitely a source of errors in our prediction.\\

Furthermore, an analysis of the individual time-series and the corresponding scales of the real-world data set in patients  has shown that the underlying Markov Chains for the markers are not strictly left-right\footnote{Left-Right HMMs model strictly degrading states, meaning once a state transition from $A$ to $B$ has occurred, it's impossible to reverse this transition. $A$ will never be reached again.}. To put it in other words, the difference in, e.g., the ADL score, a measure of the subject's ability to manage activities of daily living, between the last and first measurement is not strictly positive. In some cases, we can observe that a patient reduces their score over time, instead of displaying a degenerating performance. There are mainly potential reason for such an unexpected improvement over time. One explanation can be an initially present comorbidity causing an impairment beyond the ataxia disorder, e.g. a broken arm. Another, probably more common, explanation might be a rater depended on variance. 
In the case of predicting the hidden markers ``ADL Score'', ``INAS Score'' or ``SARA Score'' the granularity of the discretization of the continuous scales deeply impacts the prediction score. For the given data, the granularity (i.e. the dtype linspace defined in the configuration) was kept the same for all measurements. In practice, the user would be able to change the number of categories, the continuous values that each scale can fall into, to reduce the number of misclassifications. Simple trail-and-error experiment runs showed that the performance can be improved to an $F_1$-score of about $0.5$ by reducing the number of categories from about 20 down to 5. Surprisingly, doing this has a negative effect on the prediction results for the marker ``diagnosis'', as it seems an over-simplification does weaken the precision and recall of our prediction.

\subsection{Strengths, Weaknesses and Improvements}
Since the model can be seen as a mixture of HMMs, its capabilities to capture deep complexity and intricate relationships within data is limited. As the model can use multiple markers for prediction, the general stability of a prediction is traded for a lack of attention to details in the data. Since we weigh the influence the markers have on the prediction separately, small but maybe very critical changes in a single marker might be overpowered by the sheer number of markers our model has to respect. The only way to combat this problem right now would be to assign a higher weight to critical markers, in order to let small changes in states for an important marker have a higher impact on the final prediction. 
However, a data-driven solution that includes an automated identification of helpful weightings would be desirable. This could be realized by employing an optimization technique such as gradient based optimization or a probabilistic approach such as a genetic algorithm. 

Generally, Hidden Markov Models do model, as their name suggests, hidden states, including the initial state probabilities, transitions and emission signal probabilities. The usual case is that these three parameters are learned or approximated, as already stated in Section \ref{section:model_overview}. However, in this work, we have assumed an observable state to be the hidden state. This was necessary in order to maintain the flexibility and generality of the model, although of course this is not a typical use-case of HMMs. Subsequently, we were able to omit the step of training an HMM on data, since the training data already provided us with precise measurements. Thus, there was no need to approximate which parameters might have led to the displayed data -- the parameters could be extracted right away. In the future, we will think about using the extracted probabilities as a starting point for the optimization. As the training of HMMs is very sensitive to the choice of the initial guess for the parameters, see \cite{rabiner1989tutorial}, another optimization approach would be to 1) extract the probabilities for $\mathcal{A}, \mathcal{B}, \pi$, 2) possibly apply smoothing to the parameters and 3) then let an optimization algorithm find a local optimum for said parameters. This was presented by \cite{chau1997optimization}, who optimized their HMM with a genetic algorithm.

\subsection{Model Capabilities and Usecases}

During the design of the pipeline, we paid special attention to usability. The prediction target and result can be altered by specifying the prediction-layers and hidden marker, but also by specifying the weight functions for the layers and their respective markers, as well as the partition of the markers into layers in the first place. Domain experts are free to adjust these parameters within the constraints of the proposed abstract syntax. Apart from configuring the models parameters, domain experts are presented with multiple prediction capabilities:

1. \emph{Posteriors for states of a hidden marker}: The prediction tool is able to predict and plot posteriors for the states of a hidden marker $M_{\mathcal{H}}$. For a given observation sequence, the model predicts $\alpha_t(i)$ for every state $i$ and every time step $t$ of the observation sequence. This enables the user to produce a simple visualization of how likely the model thinks a certain hidden state for the given time step of the observation. A domain expert can see at a glance in which direction the states of $M_{\mathcal{H}}$ evolve.

2. \emph{Approximation of the stationary distribution}: Additionally, the model can give a rough estimate of the ``future'' posteriors of the states of $M_{\mathcal{H}}$ beyond the given observation sequence. In other words, it can give an estimate about how the distribution over the probability for the states of $M_{\mathcal{H}}$ will evolve over future time steps. It should be added that this estimate is a visualization of the approximation of the stationary distribution of the state transition matrix $\mathcal{A}$ of the underlying Hidden Markov Model.

3. \emph{Optimal state sequence prediction}: The most valuable prediction capability might be the prediction of optimal state sequences given an observation sequence. This allows for ``double-checking'' already found observation sequences, and might be used for data augmentation in the case of missing data or that a malfunctioning sensor gives back erroneous measurements. Predicting an optimal sequence of states for a hidden marker ``diagnosis'' might be a help to a medical professional, who wants to verify their given diagnosis.

4. \emph{Extraction of model parameters}: Finally, a user can extract the model parameters themselves, as they can give us critical information about the general transition probabilities of a model. A domain expert might ask about a rough estimate of the probability for state transitions, which they could immediately obtain by extracting the state transition probability matrix $\mathcal{A}$ from the model.

\section{Conclusions}

Biomedical data often contains longitudinal data, for example biomedical information on disease progress. An important goal is to infer the unknown solely from observation.
Hidden Markov Models (HMMs) have been successfully applied to the processing of possibly noisy continuous signals. Here, we presented a novel approach based on multivariate time-series of categorically distributed data.

We provided a prediction pipeline system, which processes data paired with a configuration file, enabling to construct, validate and query a fully parameterized HMM-based model. It has been conceptualized to be highly customizable and accessible both to computer scientists and practitioners from other disciplines, for example biomedical research. 
In addition, we provided a theoretical and practical framework for multivariate time-series inference based on HMMs  that included constructing multiple HMMs to predict another observable variable. Our analysis was carried out on random data, but also on biomedical data based on Spinocerebellar ataxia type 3 disease.

The implementation of the HMM framework is publicly available and can be easily configured and adapted for further experiments. 
We could show that our proposed approach shows promising results when tested on random data, and in particular good results on real world application data. However, we also presented a detailed discussion and how the approach could be augmented to improve results. It is obvious that the selected input data must meet certain criteria, especially with respect to the quantity of available longitudinal time points, to allow reasonable predictions. But even in a relatively small data set of rare diseases, we could demonstrate the feasibility of our approach. The presented framework has to be proven on further different biomedical time-series in order to fully estimate its potential and moreover how the suggested augmentations might improve the overall prediction.


\appendix

\section*{Funding}
This work was supported by a postdoc fellowship of the German Academic Exchange Service (DAAD), granted to RR.

JF receives funding of the National Ataxia Foundation (NAF) and as a fellow of the Hertie Network of Excellence in Clinical Neuroscience.

 This publication is an outcome of ESMI, an EU Joint Programme - Neurodegenerative Disease Research (JPND) project (see \url{www.jpnd.eu}). The project is supported through the following funding organisations under the aegis of JPND: Germany, Federal Ministry of Education and Research (BMBF; funding codes 01ED1602A/B); Netherlands, The Netherlands Organisation for Health Research and Development; Portugal, Foundation for Science and Technology and Regional Fund for Science and Technology of the Azores; United Kingdom, Medical Research Council. This project has received funding from the European Union’s Horizon 2020 research and innovation programme under grant agreement No 643417.


\section*{Availability of data and materials}
The pipeline and additional material is available at \url{https://github.com/rfechner/generic-hmm}. The biomedical data that support the findings of this study underlie data protection policies and are therefore not publicly available. However, they can be made available upon reasonable request with permission of the ESMI consortium (contact: Jennifer Faber, \url{Jennifer.faber@dzne.de}).







\bibliographystyle{bmc-mathphys} 





\end{document}